\documentclass[twocolumn,prd,amsmath,amssymb,superscriptaddress]{revtex4}

\usepackage{scrextend}
\usepackage{epsfig,psfrag,graphics,verbatim}
\usepackage{graphicx}
\usepackage{dcolumn}
\usepackage{bm}
\usepackage{xcolor}
\usepackage{float}
\usepackage{epsfig}
\usepackage[english]{babel}
\usepackage{amsmath} 
\usepackage{verbatim} 
\usepackage{mathrsfs}
\usepackage{amsfonts}
\usepackage{amssymb}
\usepackage{soul} 
\usepackage[normalem]{ulem}
\usepackage[latin1]{inputenc}
\usepackage{hyperref}
\usepackage{subfig}
\usepackage{color}
\usepackage{tabulary}
\newcolumntype{K}[1]{>{\centering\arraybackslash}p{#1}}

\newcommand{\beq}{\begin{equation}}
\newcommand{\eeq}{\end{equation}}
\newcommand{\barr}{\begin{eqnarray}}
\newcommand{\earr}{\end{eqnarray}}
\newcommand{\bea}{\begin{eqnarray*}}
\newcommand{\eea}{\end{eqnarray*}}

\begin{document}

\title{Thick shell regime in the chameleon two-body problem} 

\author{Lucila Kraiselburd}
\affiliation{Grupo de Astrof\'{\i}sica, Relatividad y Cosmolog\'{\i}a, 
        Facultad de Ciencias Astron\'{o}micas y Geof\'{\i}sicas, 
        Universidad Nacional de La Plata,
        Paseo del Bosque S/N 1900 La Plata, 
        Argentina.}
\affiliation{CONICET, Godoy Cruz 2290, 1425 Ciudad Aut\'onoma de Buenos Aires, Argentina.}        

\author{Susana Landau}
\affiliation{CONICET, Godoy Cruz 2290, 1425 Ciudad Aut\'onoma de Buenos Aires, Argentina.} 
\affiliation{Departamento de F\'{\i}sica and IFIBA, 
        Facultad de Ciencias Exactas y Naturales, 
        Universidad de Buenos Aires , 
        Ciudad Universitaria - Pab. I, 
        Buenos Aires 1428, 
        Argentina.}

\author{Marcelo Salgado}
\affiliation{Instituto de Ciencias Nucleares, Universidad Nacional Aut\'{o}noma de M\'{e}xico, A.P. 70-543, M\'{e}xico D.F. 04510, M\'{e}xico.}

\author{Daniel Sudarsky}
\affiliation{Instituto de Ciencias Nucleares, Universidad Nacional Aut\'{o}noma de M\'{e}xico, A.P. 70-543, M\'{e}xico D.F. 04510, M\'{e}xico.}
\affiliation{Department of Philosophy, New York University,  New York, NY 10003,   United States of America.}

\author{H\'{e}ctor Vucetich}
\affiliation{Grupo de Astrof\'{\i}sica, Relatividad y Cosmolog\'{\i}a, 
        Facultad de Ciencias Astron\'{o}micas y Geof\'{\i}sicas, 
        Universidad Nacional de La Plata,
        Paseo del Bosque S/N 1900 La Plata, 
        Argentina.}

\date{\today}
\begin{abstract}

  In a previous paper \cite{KLSSV18} we pointed out some shortcomings of the {\it standard approach} to chameleon theories consisting in treating the small bodies used to test the Weak Equivalence Principle (WEP) as test particles, whose presence do not modify the chameleon  field configuration. In that paper we developed an alternative method to  determine  the relevant  field  configuration which takes into account the influence  of both  test and source bodies, and computed the  chamaleon mediated force. Relying on that  analysis  we showed that  the effective acceleration of test bodies  is composition dependent even when the model is based on universal couplings. In this paper, we improve our method by using a more suitable  approximation for the effective chameleon potential in situations where the  bodies are in the  so called ``thick shell regime''. We then find  new and  more restrictive  bounds on the model's parametres  by confronting the   new theoretical predictions with the  empirical bounds  on    E\"{o}tv\"{o}s parameter comming from the Lunar Laser Ranging experiments.    
 
\end{abstract}

\maketitle

\section{INTRODUCTION}
 The realization that  a dark sector of   physics  that is  essentially  undetected    except for is  gravitational signatures  has led  physicists to  contemplate  the  existence of various  new  kinds of fields  beyond what   it is found  in the   Standard Model of Particle Physics.  Among these  are  scalar fields  considered as alternatives to a  simple cosmological constant, which is usually invoked  to account for the late time  accelerated  expansion  of the universe.  One obstacle  that such proposals need to face is that if these fields    are  cosmologically relevant they   would naturally tend  to  generate  long  range forces among material bodies that    would  generically led to effective  violations of the  ``universality of free  fall", making them empirically unviable.     
The scalar field model proposed by Khoury \& Weltman \cite{KW04,KW04b} (dubbed chameleon) is an alternative
which seems to  evade  such problem.  In this model a scalar field $\varphi$ is responsible for the late time accelerated expansion of the Universe. This scalar field couples nonminimally 
and nonuniversally to the fundamental matter fields, and minimally to the curvature (in the Einstein frame),  and thus the model leads, {\it in principle}, to  effective  violations of the  universality of free fall (we will refer to this feature as  an  effective violation of the  
Einstein's Weak Equivalence Principle (WEP)
despite the fact that strictly  speaking,  the theory  is  in complete accord  with general covariance).
The key ingredient of this model,  which  nonetheless, makes it {\it in principle viable},  is its  seeming ability   to  evade the stringent experimental  bounds on the violation of the 
WEP. This  is tied  to the fact that the effective mass of the chameleon depends on the density of the medium where the field propagates and therefore the model 
develops screening or {\it thin shell} effects that can in principle suppress the experimental violations of the WEP.

The model  has been previously  scrutinized 
by several authors\cite{Brax04,MS07,Hui2009,Brax2010,BB11,K13}.  Those  works   indicate    that  the  model  deals  successfully with the stringent  bounds on the experimental violation of the WEP that are imposed by several observations within the Solar System, including the 
laboratory experiments performed on Earth, like the Eot-Wash torsion balance \cite{Schlamminger2008}, and the Lunar Laser Ranging experiment\cite{LLR}. 
The  point is  that the  chameleon model predicts that in regions of high-density contrast a {\it screening effect} 
accounts for suppressing the propagation of the field due to the presence of {\it thin-shell} effects,  
while in environments with a low-density contrast the chameleon field is enhanced due to the presence of {\it thick-shell} (or ``unscreened'') effects. 
Thus, the {\it thin-shell} suppression allows the chameleon model to evade the bounds on the WEP for certain values of its parameters.

In most of the previous works found in the literature the chameleon field is studied   as a  
 single-body problem, with an  environment using suitable ``linear'' approximations for the chameleon effective potential. 
We call this method, the {\it standard approach}. These approximations are in good agreement with the numerical solutions to the full nonlinear one-body problem, but  they explicitly 
neglect the effects on the field by the test bodies themselves. Such  effects   might  be   quite important  given the 
inherent nonlinearity of the chameleon  equation. Once the field profile is obtained, the next step is to  estimate the force  acting on a test body, 
which, under the standard approach, is computed by  considering the gradient  of the  scalar field previously  obtained,  and  making  use of a heuristic   argument  (rather than a direct  methodical  calculation)  that  relies  on   approximations associated  with  the  size  and   composition of  the particular  test  body  under consideration to determine its  {\it effective  charge}, and thus  proceed   to estimate   the force  as a product of  such   charge   with the previously computed  field's gradient.  Again,  the nonlinearity of the chameleon  equation   indicates  that   such  an  approach might  not produce   completely accurate predictions.


The {\it standard approach}  relies heavily on the consideration of  two regimes: a) the {\it thin shell regime} where the field settles near the two minima of the effective potential (inside and outside 
the body) except within a thin  region  (or shell) inside the body near its surface, where the field interpolates between the two minima; b) the 
{\it thick shell regime} where the field inside the body is close to the minima of the effective potential associated with the environment. 
In this regime the effective potential is dominated by the linear contribution that is proportional to the density of the body to which the chameleon couples. 
As we emphasized above, in the {\it thin shell regime} violations of the WEP are supposed to be strongly suppressed, but not in the {\it thick shell regime}.  However, we  must keep in mind  that  those   approximations   would   be   100$\%$   accurate only in  some  unphysical limits (infinite or  zero  size  bodies), and thus,   depending on  the  accuracy one is  interested in,  deviations from the  estimates   resulting from {\it standard approach} might  occur.

 The analytic treatments  in the two  regimes require different approximations for the effective chameleon potential. 
According to the {\it standard approach}, in the {\it thick shell regime} 
the chameleon force on a test body is not suppressed and is composition dependent when the chameleon coupling to matter $\beta$ is {\it not} universal. Given   that  in the {\it standard approach} the test bodies are treated as 
point-like particles the differential acceleration between the test bodies is proportional to the difference between their  corresponding  chameleon couplings $|\beta_1-\beta_2|$.  The constant of proportionality is related to the bulk properties of the large body, such as  its mass $M_c$, size $R_c$, coupling $\beta$ and most importantly, it is related to its {\it thin-(thick)} shell parameter $\Delta R_c/R_c$. Thus, provided $\Delta R_c/R_c \ll 1$, the 
difference $|\beta_1-\beta_2|$, which (barring fine tunings) is expected to be of order one, is suppressed by the small factor
$\Delta R_c/R_c$ in  such a 
way that the E\"otvos parameter $\eta= 2|a_1-a_2|/(a_1+a_2)$ satisfies  the bound $\eta\leq 10^{-13}$ arising from the E\"ot-Wash 
experiments~\cite{Schlamminger2008}
\footnote{The philosophy behind the 
chameleon model is that the difference between the couplings $|\beta_1-\beta_2|\sim 1$ in order to avoid unphysical fine tunings. That is, 
if the {\it thin-shell} suppression were not present then in order to satisfy the observational bounds on $\eta$ one would require to 
accommodate this difference to be $|\beta_1-\beta_2|\lesssim 10^{-13}$ for every combination of two test bodies used in an experiment. 
Clearly such  requirement would make the chameleon model rather  unappealing as an alternative to account for the current speedup of the Universe's  expansion rate.}. Conversely, if $\Delta R_c/R_c\sim 1$ the chameleon force becomes 
macroscopically significant  and would have been detected by  existing  experiments. Since this is not the case, the model requires {\it a fortiori} the 
{\it thin shell} effects to avoid the observational constraints. Only then the model can be considered as a serious candidate for explaining the current accelerated 
expansion of the Universe.

In the {\it standard approach}, test bodies are treated as point-like particles and therefore the chameleon field (from which the force is derived) is computed considering just the source body. If one did not account in any way for the fact that actual test bodies are  not truly point-like, it is clear that (assuming a universal coupling scenario) the resulting estimates for the chameleon mediated force would turn out to be composition independent leading to identically vanishing estimates for the E\"otv\"os parameters. In order to deal with the test bodies' finite size, the {\it standard approach} prescribes the introduction of a ``form factor" correcting the computed force of a point-particle \cite{KW04b}. This cannot be considered as a truly reliable and satisfactory treatment of the issue, and should be regarded instead as just a well motivated method for extracting an estimate of the force. It is clear that what is required is a method that permits the explicit computation of the force that takes into account from the start the fact that actual experiments are  preformed with finite size test bodies.

 Thus,  in order to obtain  a more realistic  estimate  for the chameleon force between a source or ``large" body and a small ``test'' body,   an   
 analysis based on  two-body  characterization of  problem  is required. That  is,   we need   a   treatment  in   which both the effects of the large and   small bodies are taken into account when solving for the chameleon field. 
Under the two-body treatment the resulting chameleon force should generate directly  and automatically the {\it thin-(thick) shell} pre-factors associated with each of the 
two bodies (the large and the small ones) and also exhibit  any dependence of the force on   the corresponding   values of  $\beta$'s and other possible characteristics  of the 
bodies. Moreover, within the two-body 
treatment one should be able to clarify   whether under the assumption of a  universal-coupling  (the one that is considered  in most of the  previous papers) 
the parameter $\eta$ really vanishes or not and, if not, what is the remaining composition dependence that arises due to the  actual extended  nature of all  objects involved. 
This is precisely what we have  set to achieve  in a previous analysis~\cite{KLSSV18} by solving the chameleon field equation   in  the presence of two bodies. This 
involves solving in principle  a  highly nonlinear elliptic equation with complicated boundary conditions. We limited our analysis 
in several aspects in order to make the problem manageable: 
1) The effective chameleon  potential was approximated quadratically around their minima, one for each medium (the bodies and the environment that
surrounds them), 2) The bodies were assumed to be of constant density; 3) The  effects of gravitation on the  chameleon  field  equations  were ignored; 4) 
The large and the test bodies were assumed to be perfectly spherical; 5) The self-gravitation of the test body was neglected; 6) In 
setups that includes metal encasings around the test bodies the encasing was modeled crudely by a concentric shell of high density.
 Among these limitations, the first one was clearly unsuitable in situations where the chameleon does not ``penetrate'' deep into the effective 
potential associated with interior of the bodies. This situation corresponds to a body with a {\it thick shell} and, in that case, the field 
inside the body is far from its corresponding minimum, but very close to the effective minimum associated with the environment. 
Then, the field interpolates between the environment's minimum $\varphi^{\rm out}_{\rm min}$ 
(at spatial infinity) and the value at the center of the body which is very close to $\varphi^{\rm out}_{\rm min}$. Clearly the quadratic approximation 
we used previously breaks down in the situations where the  {\it thick shell} condition applies. The goal of this paper is to
overcome this limitation and to better approximate the potential inside 
the bodies in the appropriate situations. Since the potential in this regime is dominated by the term $\beta \rho \varphi/M_{pl}$, taking $\rho=$constant, we are led to a linear approximation. In the {\it standard approach} and  for the single body problem this approximation has been  proven to be a very good one when compared with the 
numerical solution to the nonlinear problem \cite{KW04,KW04b}. In our  methodology we do not  attempt to compare with a numerical solution to the full nonlinear 
problem  simply  because a  numerical treatment of the two-body represents a  rather complicated task. Instead we use a simple criterion based on an energy minimization argument to determine  among various treatments  which  one offers the best
accuracy for possible approximate solutions  of  the chameleon equation. In~\cite{KLSSV18} we computed and compared this energy 
for the two-body and for the {\it standard} (one body) approaches and looked for the minimum of the two 
energies. 
We concluded that 
our two-body approach provides better results than the {\it standard approach} whenever the {\it thin-shell regimes} are present in the two bodies. 
However, when the large body had a {\it thick-shell}, the {\it standard approach} turned  out to be a better suited one. As remarked above, in this paper we 
improve our two-body calculation and implement the linear approximation for the effective potential in the {\it thick-shell regime}. 
By comparing the energy of the improved solution with that of the single-body problem we find that this time 
the two-body treatment is better. Moreover, we 
show that the resulting force remains composition dependent even when universal couplings are assumed, a feature that was already present 
in our previous analysis~\cite{KLSSV18}.

The article is organized as follows; in Section \ref{sec:modelo} we present the details of the  chameleon model 
and summarize the {\it standard (single body) approach}. We also present the solution for the two body problem (obtained in our previous paper\cite{KLSSV18}) and an improved solution for the same problem with the approximation for the chameleon field's effective potential which is appropriate for the {\it thick shell} situation. In Sec.~\ref{sec:energy} we briefly review the energy criterion proposed in Ref.\cite{KLSSV18} an apply it to the situation at hand. In Sect. \ref{sec:force} and \ref{WEPP}, we analyze the chameleon force between two bodies,  compute the theoretical predictions for the E\"otv\"os parameter 
and, as an example confront them with one specific experimental setup, the Lunar Laser Ranging (LLR). Finally, in Section \ref{conclu} we present our conclusions.

\section{CHAMELEON MODEL}
\label{sec:modelo}
The chameleon model involves  a scalar-field $\varphi$ that couples minimally to gravity via a fiducial metric $g_{\mu\nu}$ (in the Einstein 
frame), but nonminimally and nonuniversally to the matter sector. The total action is
\begin{eqnarray}
\label{action}
S_T &=& \int d^4x \sqrt{-g} \left[ \frac{M_{pl}^2}{2} R - \frac{1}{2}g^{\mu\nu}(\nabla_\mu\varphi)(\nabla_\nu\varphi)- V(\varphi)\right]\nonumber\\
&-& \int d^4x L_m \left(\Psi_m^{(i)}, g_{\mu \nu}^{(i)}\right), 
\end{eqnarray}
where $M_{pl}=1/{\sqrt{8\pi G}}$ and $R$ are the reduced Planck mass and  the Ricci scalar associated with $g_{\mu\nu}$ respectively. Each specie $i$ of the matter fields $\Psi_m^{(i)}$ couples {\it minimally} to a metric $g_{\mu\nu}^{(i)}$ which is related to the Einstein-frame metric $g_{\mu\nu}$ by a  conformal factor $g_{\mu\nu}^{(i)} = \exp{\left[\frac{2 \beta_i \varphi}{M_{pl}}\right]} g_{\mu\nu}$, being  $\beta_i$  the corresponding dimensionless coupling constant between each specie of matter field 
$\Psi_m^{(i)}$ and the chameleon field.  In order to simplify the calculations we will focus only on the case of  a universal coupling $\beta_i= \beta$ in the analysis for the chameleon force between a source body and a test body, unless otherwise explicitly stated. As we will see, even in  this scenario,  the resulting force is composition dependent which  can be   understood  by noting that   variation of the density imply that  bodies of the  same  mass  have  different  size  which lead to  different  shapes for the  chamaleon field, and thus different exerted forces. 

 Like in previous analyses~\cite{KW04,KW04b,MS07,K13}, the model is characterized by a fundamental potential of runaway type, specifically $V(\varphi)= \lambda M^{4+n} \varphi^{-n}$
where $M$ is a constant that has units of mass, $n$ is an  positive or negative integer number, and  for convenience, and   following standard practice, $\lambda=1$ for all values of $n$ except when $n=-4$ when $\lambda=\frac{1}{4!}$.

The energy-momentum tensor (EMT) for each matter component, $T^{m\vspace{5mm}(i)}_{\mu\nu}$, is related with the EMT associated with the Einstein via $T^{m\vspace{5mm}(i)}_{\mu\nu}=\Big(2/\sqrt{-g^{(i)}}\Big)\delta L_m /\delta g^{\mu\nu}_{(i)}=\exp\left[\frac{-2\beta\varphi}{M_{pl}}\right]T^{m}_{\mu\nu}$. Then, the relationship between the traces of both EMT's  is given by $T^{m\vspace{5mm}(i)}=g^{\mu\nu}_{(i)}T^{m\vspace{5mm}(i)}_{\mu\nu}=\exp\left[\frac{-4\beta\varphi}{M_{pl}}\right]g^{\mu\nu}T^{m}_{\mu\nu}=\exp\left[\frac{-4\beta_\varphi}{M_{pl}}\right]T^{m}$; and a perfect-fluid and a nonrelativistic matter description is assumed for $T^{m}_{\mu\nu}$ ($T^m\approx-\rho$). From Eq.~(\ref{action}), 
\begin{equation}
\label{mov}
\Box \varphi = \frac{\partial V_{\rm eff}}{\partial \varphi},
\end{equation}
where 
\begin{equation}
V_{\rm eff}=V(\varphi) + \rho \beta \varphi /M_{pl}
\label{pot}
\end{equation}
 represents the effective potential 
 for each medium of density $\rho$ (the bodies 
and the environment) which has a minimum if $\beta > 0$. Both, the value of the field at the minimum of the effective potential 
$V_{\rm eff}$ ($\varphi_{\rm min}$) and the mass of the field   (the second derivative at  that  minimum)  $\mu_{\rm min}^2=\partial^2_{\varphi\varphi} V_{\rm eff}(\varphi_{\rm min})$ 
depend on the density $\rho$.  More  specifically  
$\varphi_{\rm min}$  decreases and $\mu_{\rm min}$ increases with the density.

\subsection{Standard approach}
Let us now consider the {\it standard approach} used to calculate the chameleon field of a single body, which was developed by several authors 
in the past~\cite{KW04,KW04b,Waterhouse06,TT08,K13}. This  analysis is restricted to 
the case where the body is considered to be static with respect to its environment and is taken to be  spherically symmetric with radius $R_c$, and with homogeneous density $\rho_{\rm in}$. Thus its mass 
is simply $M_c = 4\pi\rho_{\rm in}R^3_c /3$. Furthermore, the body is immersed in a environment of homogeneous density $\rho_{\rm out}$. 
Ignoring the backreaction of the metric Eq.~(\ref{mov}) reads
\begin{equation}
\frac{d^2\varphi}{dr^2}+\frac{2}{r}\frac{d\varphi}{dr}=V_{,\varphi}+ \frac{\beta}{M_{pl}}\rho(r).
\end{equation}  
Inside the body with density $\rho_{\rm in}$, the value of $\varphi$ at its minimum is denoted by 
$\varphi^{\rm in}_{\rm min}$, and $\mu_{\rm in}$ will denote its effective mass, 
while  $\varphi_{\infty}$ and $\mu_{\rm out}$ will denote the corresponding values associated with the environment of density $\rho_{\rm out}$ 
(i.e. outside the body). The boundary (regularity) conditions required to solve the chameleon equation are: i) $\varphi$ 
and its derivatives are bounded at the origin (for instance, $d\varphi/dr = 0$ at $r = 0$), ii) the force produced by $\varphi$ on a test particle vanishes at infinity ($\varphi\to\varphi_{\infty}$ as $r\to\infty$), and iii) $\varphi$ and  $d\varphi/dr$ should be continuous, in particular at 
the boundary between the body and its environment.

It turns out that well inside ``large'' objects with $\rho_{\rm in} \gg \rho_{\rm out}$,  in  a region $0\leq r\leq R_c-\Delta R_c$ the field is   $\varphi\approx\varphi^{\rm in}_{\rm min}$,  and it is only  within  a thin shell of thickness $\Delta R_c$
near the surface of the body, i.e. in the radial domain  $R_c-\Delta R_c\leq r \leq R_c$,  that the  behavior of  the field becomes nontrivial. 

Within that a {\it thin shell}, and to a very good approximation, $\varphi$ grows exponentially until it reaches its boundary 
at $r=R_c$ where it matches continuously the exterior solution. On the other hand, outside the object the field behaves in a typical  Yukawa form,
\begin{equation}
\label{phiKWout}   
\varphi(r)\approx-\Big(\frac{\beta}{4\pi M_{pl}}\Big)\Big(\frac{3\Delta R_c}{R_c}\Big)\frac{M_ce^{-\mu_{\rm out}(r-R_C)}}{r}+\varphi_{\infty},
\end{equation}  
where
\begin{equation}
\label{thinshell}
\frac{\Delta R_c}{R_c}\approx\frac{\varphi_{\infty}-\varphi_{\rm in}}{6\beta M_{pl}\Phi_N},
\end{equation}
and $\Phi_N=M_c/8\pi M^2_{pl}R_c$ is the Newtonian potential of the body. The {\it thin-shell condition} corresponds to $\Delta R_c/R_c\ll1$ (details leading to Eq.~(\ref{thinshell}) can be found in \cite{KW04,Waterhouse06,TT08}). 
On the other hand, when the body has a {\it thick shell}, i.e., $\Delta R_c/R_c\gtrsim1$, the value of the chameleon field does not change significantly inside and outside the body, with value $\varphi \approx \varphi_{\infty}$. This situation is typical for a
small source body and the chameleon field becomes a small perturbation of the exterior solution. For instance, the exterior solution for a  body with {\it thick shell} is,
\begin{equation}
\label{phiKWoutts}   
\varphi(r)\approx-\Big(\frac{\beta}{4\pi M_{pl}}\Big)\frac{M_ce^{-\mu_{\rm out}(r-R_C)}}{r}+\varphi_{\infty}.
\end{equation}

 and the interior solution interpolates between a value $\varphi(r=0)$ and $\varphi(r=R_c)$ which is very close to $\varphi_{\infty}$ 
in all the interior. As illustrated by Eq.(\ref{phiKWout}),   the {\it thin shell} condition is  associated  with a  field $\varphi$ whose difference with $\varphi_\infty$ is largely suppressed (i.e. screened) 
outside the object, and its gradients vanish almost everywhere (except within the {\it thin shell}) leading then to a chameleon force 
that is very small and that barely depends on the composition of the body
\footnote{However, in the  {\it thick shell regime} and for nonuniversal $\beta$, the conclusion is the opposite and large violations of the WEP are to be expected.}. This is basically the summary of the understanding of the situation as  provided by  the {\it standard approach}.

\subsection{Two-body chameleon approach}
\label{KLSSVtheory}

\begin{figure}
\includegraphics[width=4.5cm,height=7.cm]{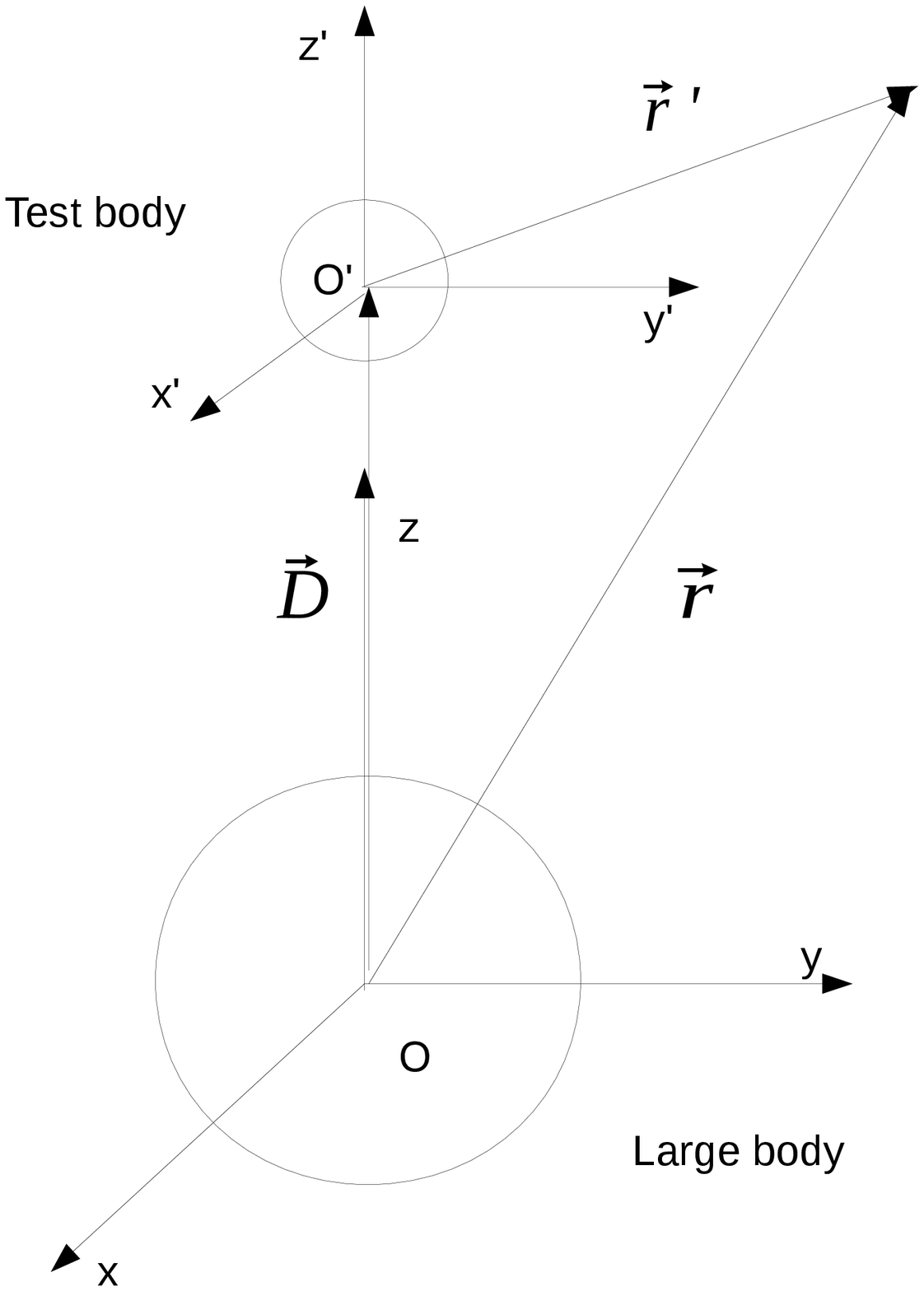}
\caption{Two body problem.}
\label{figura1}
\end{figure}

 In contrast  with  the {\it standard approach} where a test body is treated as a point-like particle, and therefore, its backreaction on the chameleon 
field is  neglected, our  method  considers    the chameleon field generated by two spherical bodies of different  and  finite size 
(see Fig.\ref{figura1})~\cite{KLSSV18}. One of the bodies corresponds to a large body, which we take as the main source 
for the gravitational field, 
and a small test body whose backreaction on the chameleon field is taken into account,  so that   the  chamaleon mediated force might be more  realistically  computed.   In  the analysis  carried out in ~\cite{KLSSV18} the effective potential was approximated quadratically around each minima (inside and outside the two bodies): 
$V_{\rm eff}(\varphi) \simeq V_{\rm eff}(\varphi_{\rm min}) +\partial_{\varphi \varphi} V_{\rm eff}(\varphi_{\rm min})[\varphi - \varphi_{\rm min}]^2/2$, 
and  the problem was studied  making use of the  axially symmetric  of the situation so the  solution for Eq.~(\ref{mov}) was described  as follows:

\begin{widetext}
\begin{equation}
\label{fullsol}
\quad \varphi=
\begin{cases}
 \varphi_{\rm in1}= \sum\limits_{lm} C_{lm}^{\rm in1} i_l(\mu_1 r) Y_{lm}(\theta,\phi)+\varphi_{1\rm min}^{\rm in} \hspace{5.25cm} 
 (0\leq r \le R_1) \\
 \varphi_{\rm out}=\sum\limits_{lm} C_{lm}^{\rm out1} k_l(\mu_{\rm out} r) Y_{lm}(\theta,\phi)+ 
 C_{lm}^{\rm out2} k_l(\mu_{\rm out} r') Y_{lm}(\theta',\phi')+\varphi_{\infty} 
 \hspace{0.7cm} (\rm exterior\,\,solution) \\
 \varphi_{\rm in2}=\sum\limits_{lm} C_{lm}^{\rm in2} i_l(\mu_2 r') Y_{lm}(\theta',\phi')+\varphi_{2\rm min}^{\rm in} \hspace{4.9cm} (0\leq r' \le R_2)
\end{cases}
\end{equation}
\end{widetext}
where $R_1$ and $R_2$ are the radii of the large and test bodies, respectively, and $i_l$ and $k_l$  are the Modified Spherical Bessel Functions (MSBF). The chameleon field is described in two coordinate systems: $(r,\theta,\phi)$ centered in body 1 (large), and $(r',\theta',\phi')$ centered in body 2 (test). Making use of the axial symmetry of the problem, the $z-$axis is established in such a way that it contains the centers of the two bodies, being $D$ the distance between them (see Fig. \ref{figura1}). Therefore, the coordinate transformation is a translation along this axis ($\vec{r}={\vec r}^{\, \prime}+D\hat{z}$) so $\theta_D=0$, and also $\phi_D$ becomes irrelevant due to the axial symmetry ($m=0$). The parameters $C_{lm}^{\rm in1}$, $C_{lm}^{\rm in2}$, $C_{lm}^{\rm out1}$ and $C_{lm}^{\rm out2}$ are determined from the boundary conditions $\varphi_{\rm in1}(R_1)=\varphi_{\rm out}(R_1)$, $\partial_r\varphi_{\rm in1}(R_1)=\partial_r\varphi'_{\rm out}(R_1)$, $\varphi_{\rm in2}(R_2)=\varphi_{\rm out}(R_2)$ and $\partial_{r'}\varphi'_{\rm in2}(R_2)=\partial_{r'}\varphi'_{\rm out}(R_2)$ by using translation coefficients $\alpha^{lm}_{vw}$ and $\alpha^{*lm}_{vw}$ (see \cite{KLSSV18} for a full description of 
the method).  Furthermore, we found  the energy functional associated with the chameleon solution $U_{\rm eff}[\varphi,\rho,D,\beta]$, which  upon extremization leads to the 
 correct equation for the static configuration. That  minimal  value, which  naturally depends on the distance between the centers of the two bodies $D$, can be  used to   compute the chameleon force 
 between the two bodies as $F_\varphi= -\partial U_{\rm eff}/\partial D$.  In the limit where the size of the test body $R_2\rightarrow 0$, i.e., in the point-particle limit, that  force reduces to the usual chameleon force $F_\varphi\sim \beta_{\rm eff} \nabla\varphi$, where $\beta_{\rm eff}$ is the chameleon coupling to the test body, and the gradient is evaluated at its position. In this case the 
solution $\varphi$ depends on the coupling $\beta$ between the chameleon and the source body, and thus, the force is proportional to $\beta \beta_{\rm eff}$. This method allows us to find the differential acceleration produced by the chameleon force on two test bodies of different composition\cite{KLSSV18}.

Our results indicate that for some choices of the free parameters of the chameleon model, the prediction for the E\"otv\"os parameter is larger 
than previous estimates obtained from the {\it standard approach} \cite{KLSSV18}.

\subsection{{\it Thick shell regime} in the two-body chameleon model}
\label{newtheory}
In our previous analysis~\cite{KLSSV18}, we did not implement the linear approximation $V_{\rm eff}^{\rm in}\sim\rho_{\rm in}\frac{\beta\varphi}{M_{pl}}$   which turns to be better suited than the quadratic approximation when bodies develop a {\it thick shell}
\footnote{We remind the reader that the linear approximation was used in the {\it standard approach}in such a regime.}. In order to determine which of these approximations is the best one, we developed a criterion using a minimization of a suitable energy 
functional. In ~\cite{KLSSV18} we found that the energy criterion favors the two body approach over the {\it standard (one body) approach} in scenarios where the large object is {\it not} in the {\it thick shell regime},   the   reverse   is true for  those  situations  in which  the bodies develop a {\it thick shell}. 
In this paper we   address  this limitation of our previous work and improve our method by 
implementing the linear approximation $V_{\rm eff}^{\rm in}\sim\rho_{\rm in}\frac{\beta\varphi}{M_{pl}}$ 
in scenarios where one or both objects are no longer in the {\it thin shell}  but in the {\it thick shell regime}.

We assume that in the {\it thick shell regime} the chameleon equation inside the body becomes,
 \begin{equation}
\label{movthick}
\nabla^2 \varphi_{\rm in} = \rho_{\rm in}\frac{\beta}{M_{pl}},
\end{equation}
while outside $\nabla^2 \varphi_{\rm out} = \mu_{\rm out}^2(\varphi_{\rm out}-\varphi_{\infty})$. According to ~\cite{Waterhouse06}, a {\it thick shell} is developed inside a body when the following condition is satisfied
\begin{equation}
\label{TSCOND}  
  \mu_{\rm in}^2\Big(\varphi_{\rm in}(0)-\varphi^{\rm in}_{\rm min}\Big)>\rho_{\rm in}\frac{\beta}{M_{pl}},
\end{equation} 
being $\varphi_{\rm in}$ the solution to Eq. (\ref{movthick}). Under those conditions, and as shown in \cite{KW04,KLSSV18}, the second term of the effective potential Eq. (\ref{pot}) dominates over the first one   and thus we take
  $V_{\rm eff}^{\rm in}\sim\rho_{\rm in}\frac{\beta\varphi}{M_{pl}}$. Conversely, when the above condition is not  satisfied the body develops a {\it thin shell}.

\begin{figure*}
\subfloat[The test body has  $thick$ $ shell$.]   
{\includegraphics[width=9.5cm,height=8.5cm]{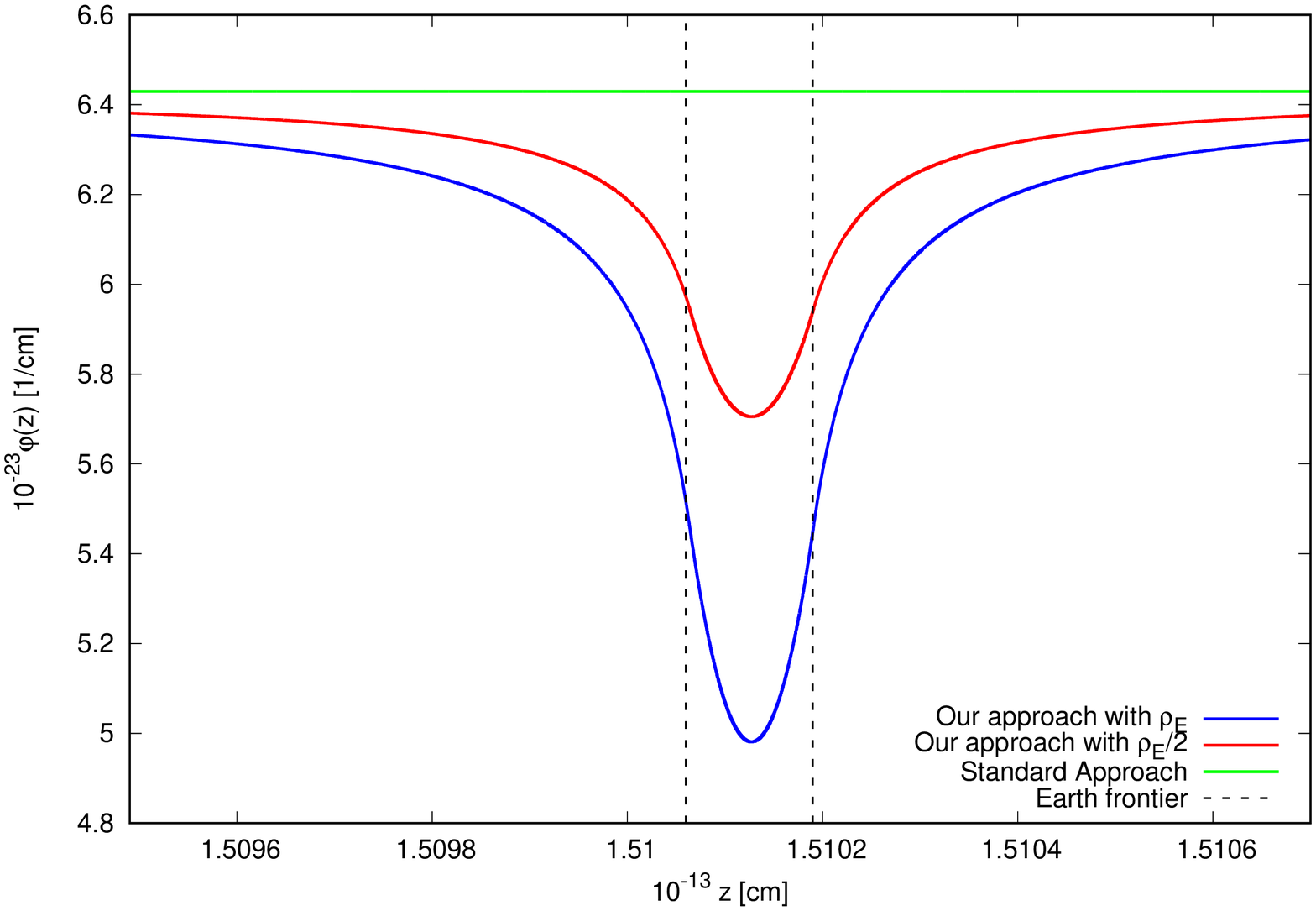}}      
\subfloat[Both bodies (large and test) have $thick$  $shell$.]
{\includegraphics[width=9.5cm,height=8.5cm]{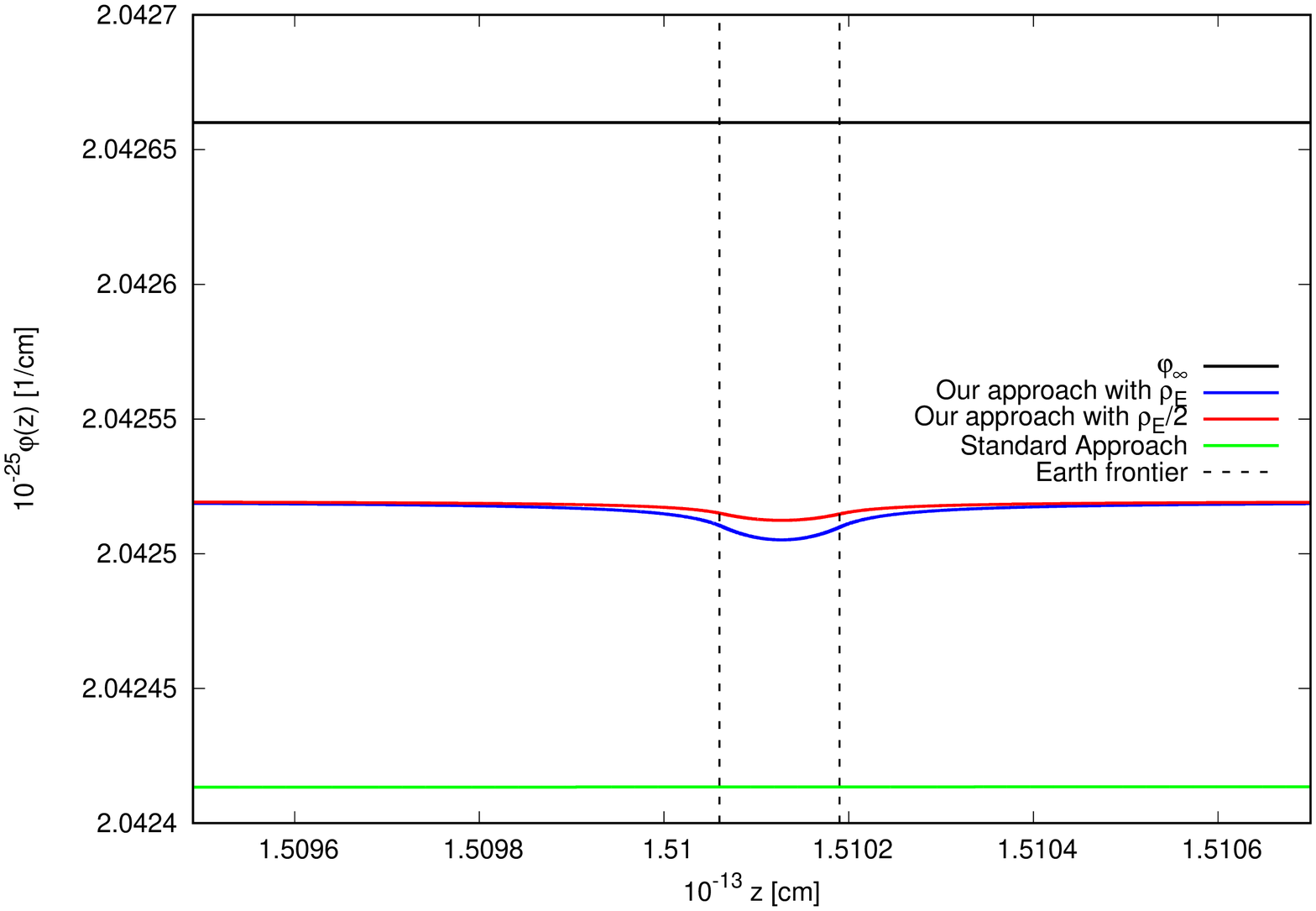}}          
\caption{The chameleon field $\varphi$ as a function of coordinate $z$ inside the test body (Earth) and in its outskirts ($z$ is the distance to the center of the large body (Sun)). The blue lines represet our approach for the Earth while the red ones depict our approach for a test body with the same radius of the Earth but with half of its density. The green lines stand for the {\it standard approach}, and the black one, the minimum value of $\varphi$ outside the bodies. Left: The Earth is in the {\it thick shell regime} but the Sun is not. Right: Both, Earth and Sun, are in the {\it thick shell regime}.}
 
\label{phirTS}
\end{figure*}

\begin{figure*}
\subfloat[The test body has  $thick$ $ shell$.]     
{\includegraphics[width=9.5cm,height=8.5cm]{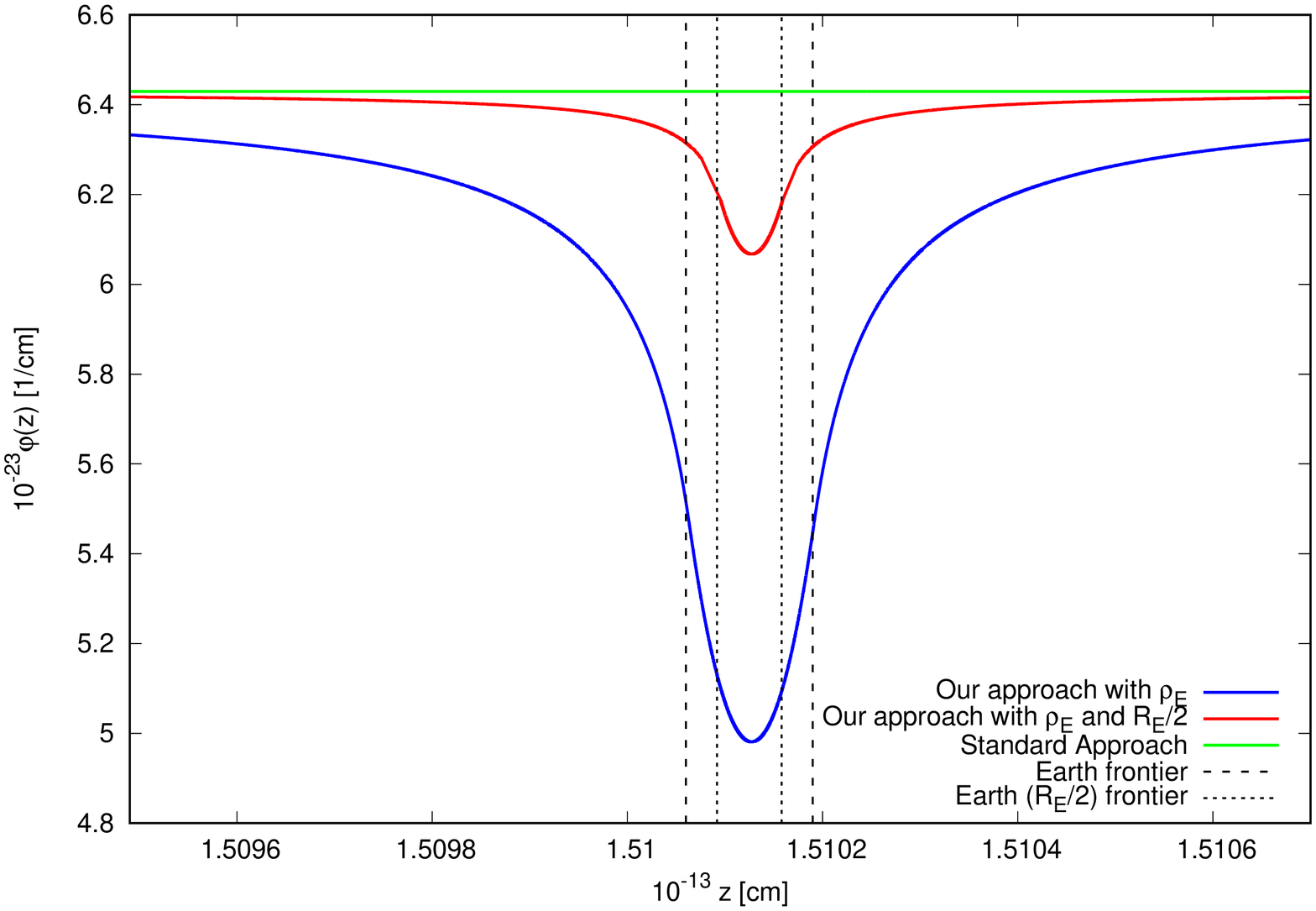}}
\subfloat[Both bodies (large and test) have $thick$  $shell$.]
{\includegraphics[width=9.5cm,height=8.5cm]{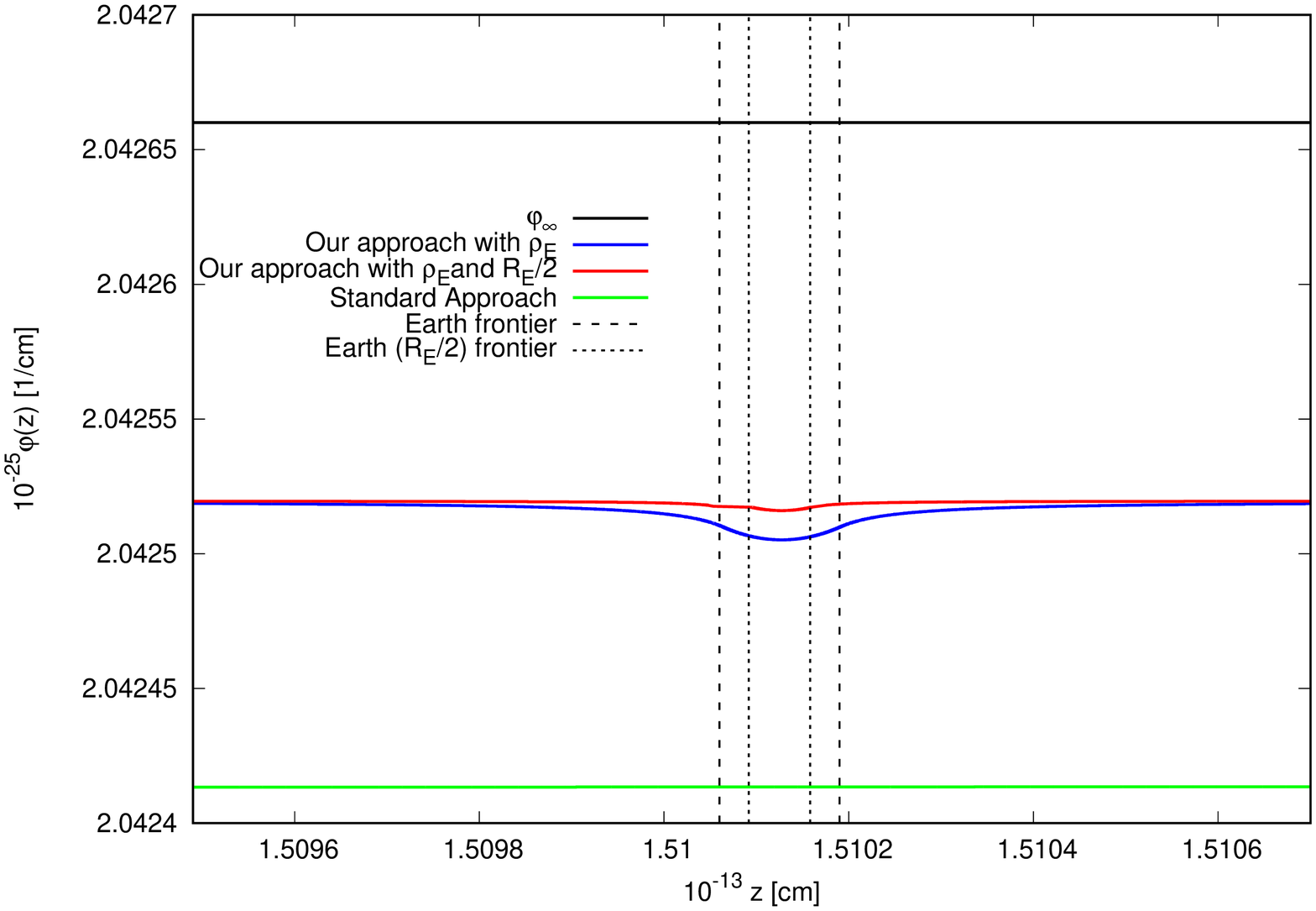}}
\caption{The chameleon field $\varphi$ as a function of coordinate $z$ inside the test body (Earth) and in its outskirts ($z$ is the distance to the center of the large body (Sun)). The blue lines represent our approach for the Earth while the red ones, our approach for a test body with the same density of the Earth but with half of its radius. The green lines stand for the {\it standard approach}, and the black one, the minimum value of $\varphi$ outside the bodies. Left: The Earth is in the {\it thick shell regime} but the Sun is not. Right: Both, Earth and Sun, are in the {\it thick shell regime}.}
\label{phirTS2} 
\end{figure*}

For instance, if the test body is the one that satisfies Eq. (\ref{TSCOND}) we expand the most general solution in
complete sets of solutions in the interior and exterior regions of the two bodies as follows:
\begin{widetext}
\begin{equation}
\label{fullsoltk1}
\quad \varphi=
\begin{cases}
 \varphi_{\rm in1}= \sum\limits_{l} C_{l}^{\rm in1} i_l(\mu_1 r) Y_{l0}(\theta,\phi)+\phi_{1\rm min}^{\rm in} \hspace{5.1cm} 
 (0 \leq r \le R_1) \\
 \varphi_{\rm out}=\sum\limits_{l} C_{l}^{\rm out1} k_l(\mu_{\rm out} r) Y_{l0}(\theta,\phi)+ 
 C_{l}^{\rm out2} k_l(\mu_{\rm out} r') Y_{l0}(\theta',\phi')+\varphi_{\infty} 
 \hspace{0.7cm} (\rm exterior\,\,solution) \\
 \varphi_{\rm in2}=\sum\limits_{l} C_{l}^{\rm in2} r'^l Y_{l0}(\theta',\phi')+ r'^2\rho_{\rm in2} \frac{\beta}{6\sqrt{\pi}M_{pl}}\hspace{4.3cm} (0 \leq r' \le R_2).
\end{cases}
\end{equation}
On the other hand, when both bodies, the large one and the test body, have a {\it thick shell} the solution can be expressed as follows:
\begin{equation}
\label{fullsoltk2}
\quad \varphi=
\begin{cases}
 \varphi_{\rm in1}= \sum\limits_{l} C_{l}^{\rm in1}  r^l Y_{l0}(\theta,\phi)+r^2\rho_{\rm in1} \frac{\beta}{6\sqrt{\pi}M_{pl}} \hspace{4.7cm} 
 (0 \leq r \le R_1) \\
 \varphi_{\rm out}=\sum\limits_{l} C_{l}^{\rm out1} k_l(\mu_{\rm out} r) Y_{l0}(\theta,\phi)+ 
 C_{l}^{\rm out2} k_l(\mu_{\rm out} r') Y_{l0}(\theta',\phi')+\varphi_{\infty} 
 \hspace{0.7cm} (\rm exterior\,\,solution) \\
 \varphi_{\rm in2}=\sum\limits_{l} C_{l}^{\rm in2} r'^l Y_{l0}(\theta',\phi')+ r'^2\rho_{\rm in2} \frac{\beta}{6\sqrt{\pi}M_{pl}}\hspace{4.3cm} (0 \leq r' \le R_2)
\end{cases}
\end{equation}
\end{widetext}
The parameters $C_{l}^{\rm in1}$, $C_{l}^{\rm in2}$, $C_{l}^{\rm out1}$ and $C_{l}^{\rm out2}$ are determined from the boundary conditions at the borders 
of the two bodies $R_1$ and $R_2$ using translation coefficients $\alpha^{lm}_{vw}$ and $\alpha^{*lm}_{vw}$ to go from one coordinate system to the other (for details see Ref.~\cite{KLSSV18}). In both cases (when the large body has {\it thin shell} and the test body does not, and when both of them are in the {\it thick shell regime}) these coefficients depend on the composition of both bodies and   on the surrounding  environment. This means that the ``chameleon'' acceleration of the test body depends on its composition. Figures \ref{phirTS} and \ref{phirTS2}   illustrate the behavior of the field $\varphi$ with respect to the test body properties in its vicinity. It can be seen that for different densities (Fig. \ref{phirTS}) and radii (Fig. \ref{phirTS2}) of the test body the field acquires different values. In Figures \ref{phirTS}a  and \ref{phirTS2}a  the large body (in this case the Sun) has a {\it thin shell} while the test body (Earth) has a {\it thick shell} with $n =\beta= 1$ and $M = 10$ eV. Meanwhile, in Figures \ref{phirTS}b  and \ref{phirTS2}b  both bodies are in the {\it thick shell regime} and $n = 1$, $\beta= 10^{-3}$ and $M = 10$ eV. Even though these differences are small, they generate a small, but  nonvanishing  dependency, of the acceleration, on the test body's composition.

\section{MINIMUM ENERGY CRITERION}
\label{sec:energy}
 We use the energy criterion proposed in~\cite{KLSSV18} to evaluate which of the three 
approximations, the {\it standard approach} \cite{KW04,Waterhouse06,TT08}, two-body approach with a quadratic effective potential~\cite{KLSSV18} or 
the two-body approach with the {\it thick shell} approximation, best characterizes the situations when one or both bodies 
develop a {\it thick shell}, and which correspond to bodies satisfying the condition~(\ref{TSCOND}). 
 The  criterion relies on the fact  that static situations, the field configuration that minimizes the energy functional of a system is also the one that extremizes the action functional 
associated with that system, and thus, corresponds to the configuration that satisfies the static  (classical) equation of motion.
Thus,  when considering  various   types of approximations to the  solution corresponding to a static field configuration,  given  the configuration of the  objects  with  which the field  interacts,  the  one   with the lowest   value  of energy functional $U[\varphi,\rho,\beta,D]$ (which is associated with the class of  test  field configurations)  offers  the  best description of  the problem.

The energy associated with the total energy-momentum tensor under the assumptions of staticity and flat space-time is $U = \int_V \left(T^{\varphi}_{00}+T^m_{00} \right) dV$. However, for the problem at hand, the energy functional that is extremized by the actual field configuration and which leads to 
Eq.~(\ref{mov}) is

\begin{eqnarray}
\label{Ueff}
U_{\rm eff} &=& \int_V \left[\frac{1}{2}(\nabla^i\varphi) (\nabla_i\varphi) + V_{\rm eff}(\varphi) \right] dV  \,\,\,.
\end{eqnarray}
We can integrate the above equation by parts and discard the surface terms that vanish at infinity. Moreover, we renormalize the resulting energy functional by 
subtracting the divergent term associated with the minimum of the effective potential of the environment, obtaining
\begin{equation}
U_{\rm eff}^{*}=\int_V \left[-\frac{1}{2}\varphi\nabla^2\varphi +V_{\rm eff}(\varphi)-V_{\rm eff}(\varphi_{\rm min}^{\rm out})   \right] dV
\,\,\,.
\label{Energydef}
\end{equation}

As a specific application to our improved approach we analyze a simple experimental scenario: the Lunar Laser Ranging (LLR) where the large body is represented by the Sun and the test bodies by the Earth and the Moon surrounded by the interstellar medium. We remind the reader that for simplicity we assume the bodies and 
the medium to be perfectly homogeneous. Thus, we compute the functional Eq.(\ref{Energydef}) 
for the three approximate methods mentioned above. As emphasized before, under the {\it standard approach}~\cite{KW04,KW04b}, 
the test body is not taken into account in the determination of the solution for $\varphi$,   and therefore a  correction factor is 
introduced   by hand in  the computation of the chameleon force in order to take into account  certain relevant  aspects of the test body. For instance, 
when the test body is considered  to have a {\it thin shell}, the correction appears in the form of a 
{\it thin shell} parameter $\Delta R/R$ of the test body, or in terms of a factor denoted $Q_B$ (see Eqs.(\ref{fkhoury1}) and (\ref{fkhoury2}) below). 
Other studies~\cite{Hui2009,Brax2010,Burrage15} introduce a correction to the solution for $\varphi$ by superposing the exterior solutions for the large 
and the test bodies but without offering a solution for the field  inside the test body (see~\cite{KLSSV18} for a thorough discussion on this issue).

\begin{figure*}
\begin{center}
\subfloat[$n=1$]
{\includegraphics[width=9.5cm,height=8.5cm]{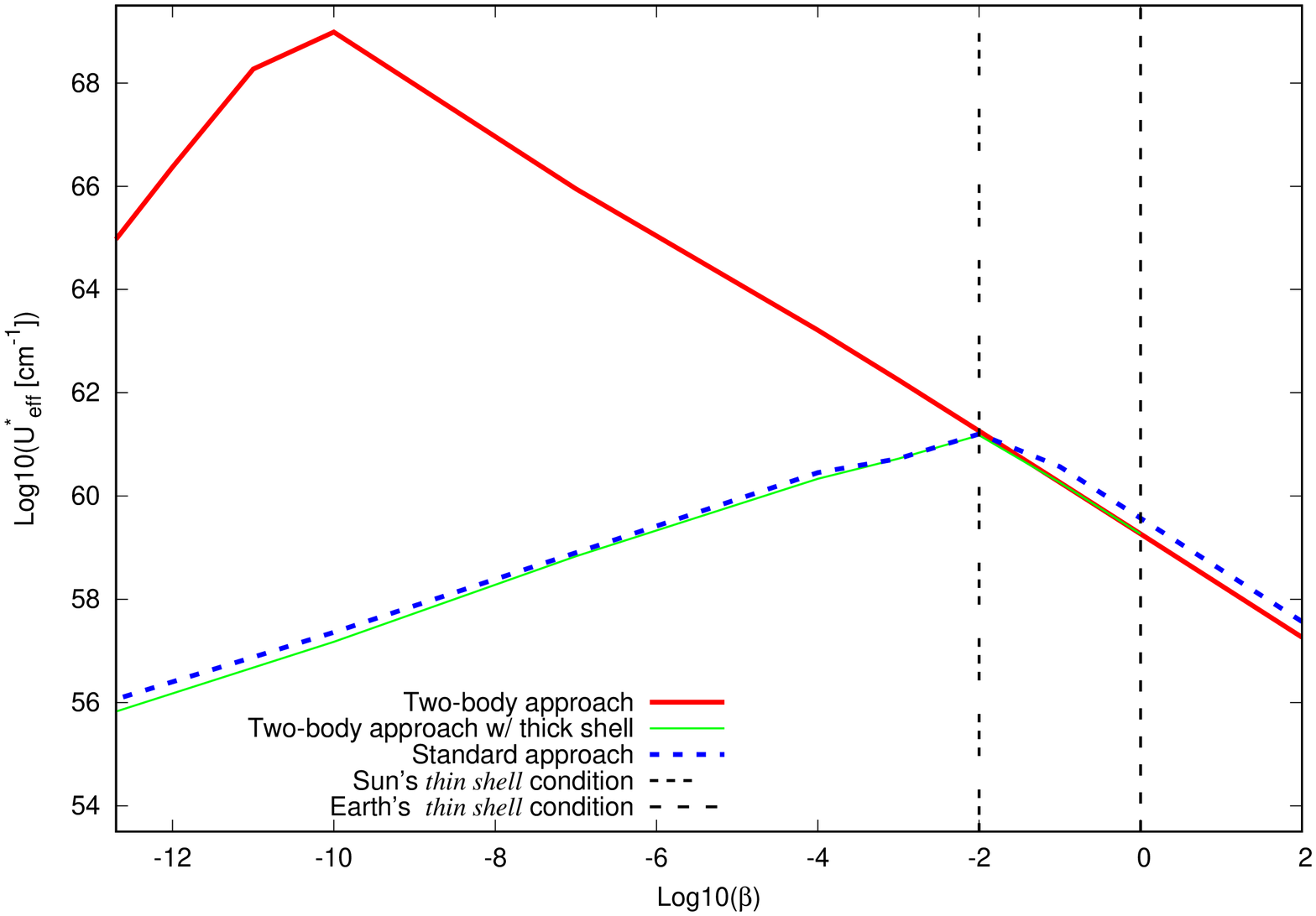}}
\subfloat[$n=1$ zoom.]
{\includegraphics[width=9.5cm,height=8.5cm]{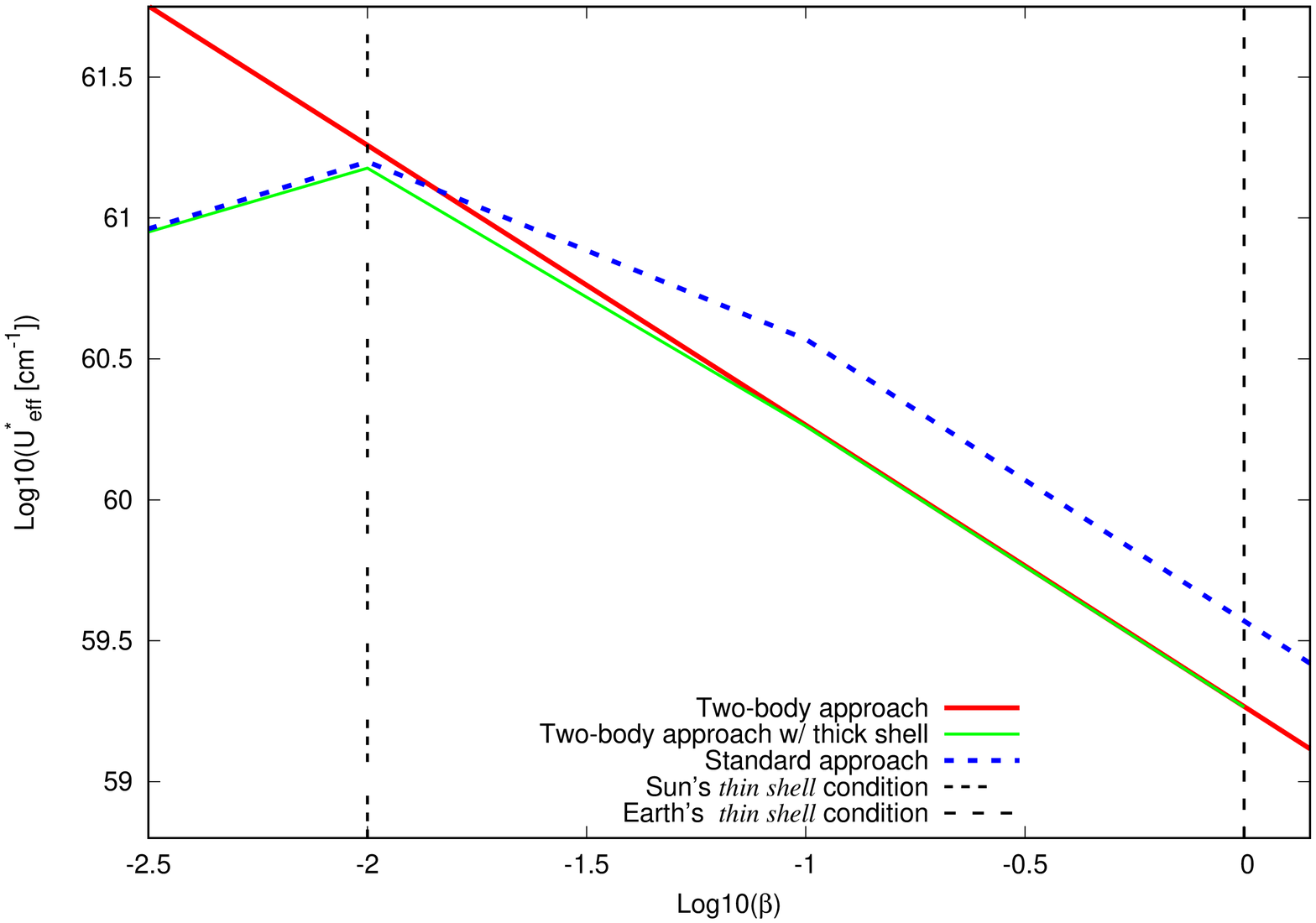}}
\caption{Energy functional computed for a setup that mimics the LLR experiment using the two-body approach (red), the two-body approach including the 
{\it thick shell} approximation (green)  and the standard one (blue) taking $M=10$ {\rm eV} and $n=1$. The Sun and the Earth correspond to the 
large and test bodies, respectively, while the environment represents the interstellar medium. The vertical dotted lines indicate the boundary of the {\it thin shell regime} for the Earth and the Sun (values of $\beta$ lower than those indicated by the vertical dotted lines point out that the bodies have a {\it thick shell}).
The right panel zooms a portion of the left panel. Notice that the energy functional is minimized for the improved two body problem (green line).}
\label{cenergy1}
\end{center}
\end{figure*}

\begin{figure*}
\begin{center}
\subfloat[$n=2$]
{\includegraphics[width=9.5cm,height=8.5cm]{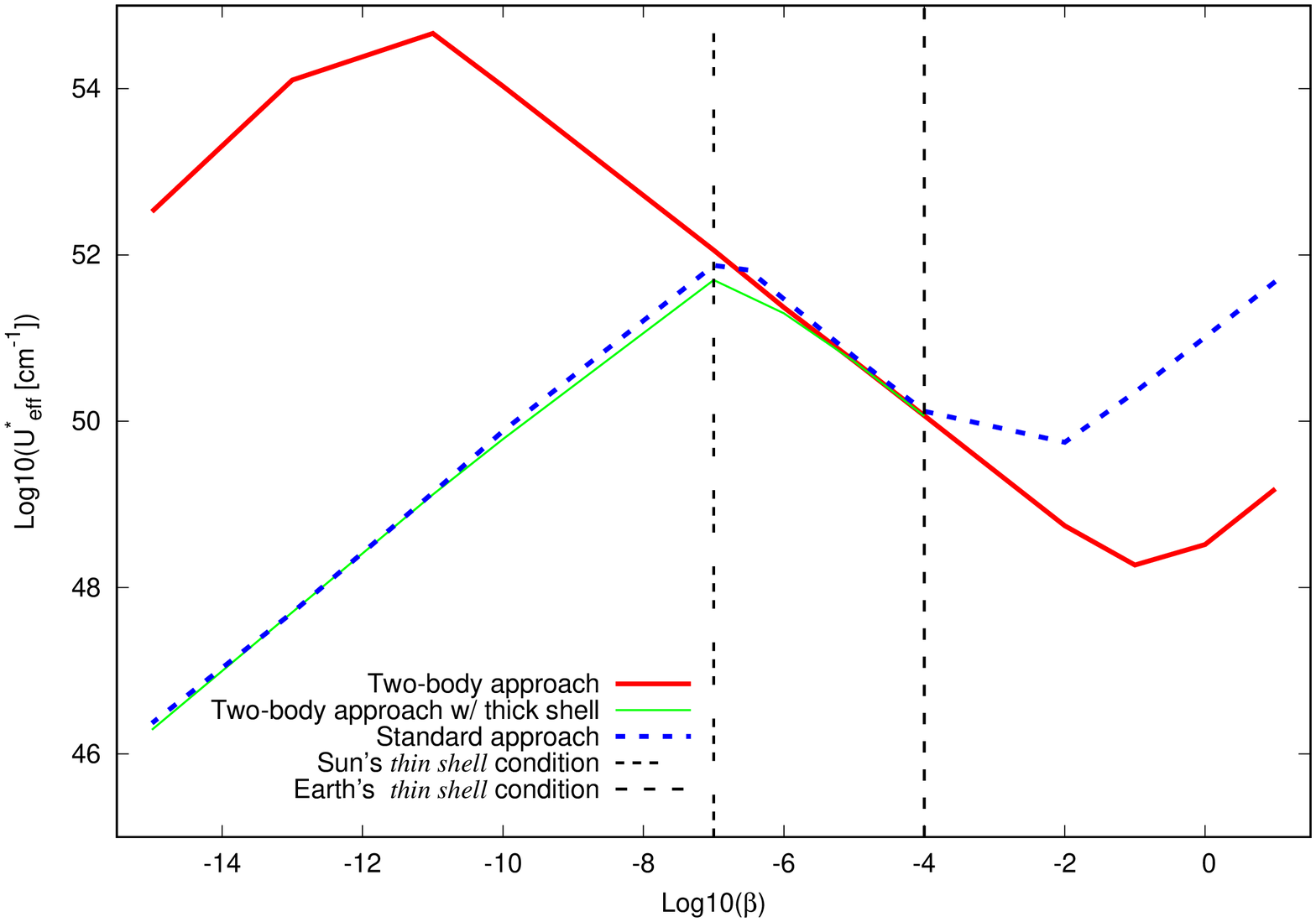}}
\subfloat[$n=2$ zoom.]
         {\includegraphics[width=9.5cm,height=8.5cm]{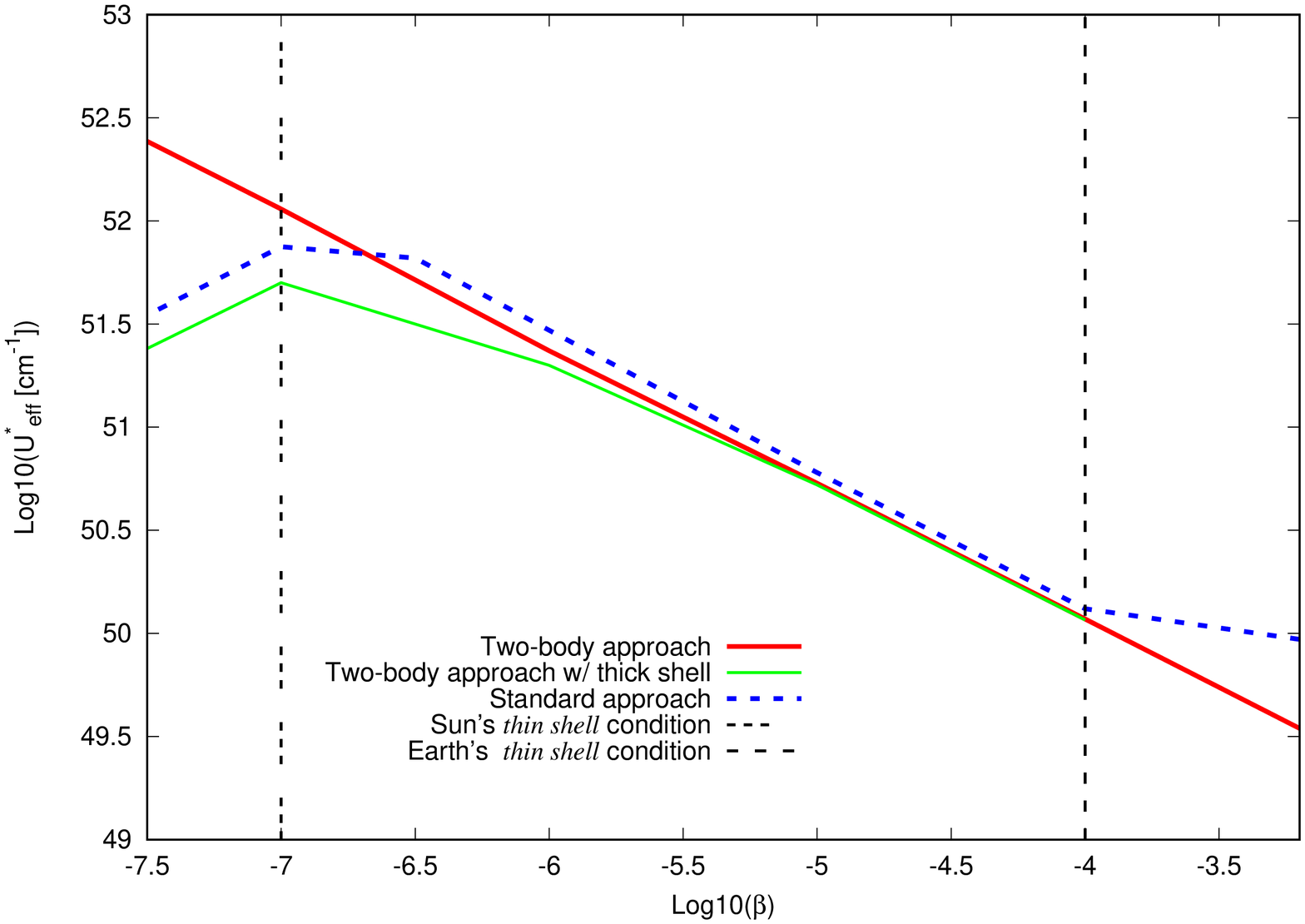}}         
\caption{Similar to Figure~\ref{cenergy1} taking $n=2$ .}
\label{cenergy2}
\end{center}
\end{figure*}

Figures \ref{cenergy1} and \ref{cenergy2} show the energy functional computed  for the different approximations 
({\it standard approach}, two-body approach, two-body approach with {\it thick shell}), taking $M=10$ {\rm eV}, $n=1,2$ and 
$\rho_{\rm Sun}=1.43 \hspace{0.1cm}  {\rm g  \hspace{0.1cm} cm}^{-3}$, $R_{\rm Sun}= 7 \times 10^8$ m, $\rho_{\rm Earth}=5.5 \hspace{0.1cm} {\rm g  \hspace{0.1cm} cm}^{-3}$, $R_{\rm Earth}=6.371 \times 10^6$ m,  $\rho_{\rm Moon}=3.34  \hspace{0.1cm}{\rm g  \hspace{0.1cm} cm}^{-3}$, $R_{\rm Moon}=1.737 \times 10^6$ m and  the interstellar medium density $\rho_{out}=10^{-24} {\rm g \,\ cm^{-3}}$. We appreciate that when the bodies are in the {\it thick shell regime} 
the energy functional is minimized when the improved two-body approach is implemented.

\section{CHAMELEON MEDIATED FORCE BETWEEN TWO SPHERICAL OBJECTS}
\label{sec:force}
Under the standard approach~\cite{KW04,Hui2009,Burrage15} the chameleon force between a source body $A$ and a test body $B$ is computed from
\begin{subequations}
\begin{equation}
F^{AB}_{\varphi}=2Q_{A}Q_{B} F_N,
\label{fkhoury1}
\end{equation}
\begin{equation}
Q_i=min\left(\beta,\frac{|\varphi_{\infty}-\varphi^{\rm in}_{i\rm min}|}{2M_{pl}\Phi_{N_{i}}}\right),
\label{fkhoury2}
\end{equation}
\end{subequations}
where $i=A,B$, $F_N$ refers to the gravitational force between the bodies, and  $\Phi_{N_{i}}=\frac{G{\cal{M}}_{i}}{R_i}$ refers to the Newtonian potential of the body $i$. When one of the two bodies has a {\it thin shell}, the chameleon force is largely 
suppressed since $Q_{A,B}=\frac{|\varphi_{\infty}-\varphi^{\rm in}_{A,B\rm min}|}{2M_{pl}\Phi_{N_{A,B}}} \ll1$. However, when the bodies have a {\it thick shell} $Q_{A,B}\sim \beta$ (assuming that all the couplings are of the same order), the acceleration on the test body due to the chameleon force turns to be independent of its composition
\footnote{However, when the coupling $\beta$
 is not universal, and in the {\it thick shell regime}, the E\"otvos parameter for two test bodies is not longer suppressed and it is proportional to the difference of the couplings of the two test bodies 
(which is of order one) and large violations to the observational bounds are expected.}.

\medskip

\subsection{Chameleon force in the two-body approach}
In our previous analysis~\cite{KLSSV18} we computed the effective chameleon force between the large and the test body
from the first principles using the effective energy functional Eq.(\ref{Ueff}) for a given configuration as follows: 
$F_{\rm z \varphi}=-\frac{\partial U_{\rm eff}}{\partial D}$, where $D$ is the distance between the center of the two bodies. 
After some simplifications, the force reads
\begin{widetext}
\begin{eqnarray}
\label{fuerzaC}
F_{\rm z\varphi} &=& F_{\rm z\varphi_{\rm in2}}+F_{\rm z\varphi_{\rm in1}}+F_{\rm z\varphi_{\rm out}}\nonumber\\
&=&\frac{\partial}{\partial D}\int_{V_2}\Big\lbrace\hat{\varphi}_{\rm in2}(\vec{r}^{\,\prime})\Big[\frac{(n+1)}{2}\frac{\rho_2\beta}{M_{pl}}\Big]-\frac{(2+n)\mu_2^2\hat{\varphi}^2_{\rm in2}(\vec{r}^{\,\prime})}{4}\Big\rbrace 
d^3{\vec r}^{\,\prime} \nonumber\\
&+& \frac{\partial}{\partial D} \int_{V_1}\Big\lbrace\hat{\varphi}_{\rm in1}(\vec{r})\Big[\frac{(n+1)}{2}\frac{\rho_1\beta}{M_{pl}}\Big]-\frac{(2+n)\mu_1^2\hat{\varphi}^2_{\rm in1}(\vec{r})}{4}\Big\rbrace 
d^3{\vec r} \nonumber\\
&+&\frac{\partial}{\partial D}\int_{V_3}\Big\lbrace\hat{\varphi}_{\rm out}(\vec{r})\Big[\frac{(n+1)}{2}\frac{\rho_{\rm out}\beta}{M_{pl}}\Big]-\frac{(2+n)\mu_{\rm out}^2\hat{\varphi}^2_{\rm out}(\vec{r})}{4}\Big\rbrace 
d^3{\vec r} \nonumber\\
&-& \frac{\partial}{\partial D}\int_{V_2}\Big\lbrace\hat{\varphi}_{\rm out}(\vec{r}^{\,\prime})\Big[\frac{(n+1)}{2}\frac{\rho_{\rm out}\beta}{M_{pl}}\Big]-\frac{(2+n)\mu_{\rm out}^2\hat{\varphi}^2_{\rm out}(\vec{r}^{\,\prime})}{4}\Big\rbrace d^3{\vec r}^{\,\prime}\;,
\end{eqnarray}
\end{widetext}
where ${\hat \varphi}= \varphi-\varphi_{\rm min}$. $V_1$ represents the region  occupied  by  the large body; $V_2$, the region  occupied  by  the test body; while $V_3$, the region outside the large body in the coordinate system centered in the large body, the last  term  compensates for the   fact that $V_3$,  includes the test body:
\begin{equation}
\quad V_3=
\begin{cases}
 R_1\leq r\leq \infty \\
 0\leq\theta\leq\pi\\
 0\leq\varphi\leq 2\pi
\end{cases}
\end{equation}

\subsection{Chameleon force in the two-body approach in the {\it thick shell regime}}

We use a similar expression for the force but by implementing the improved approximation and its solution for the chameleon field as described in 
Sec.~\ref{newtheory}. Thus, when  only one of the bodies is in the {\it thick shell regime}, for instance, the test body (body 2) 
the expressions $F_{\rm z\varphi_{\rm in1}}$ and $F_{\rm z\varphi_{\rm out}}$ remain the same as in Eq.~(\ref{fuerzaC}) 
except that the solution for $\varphi$ is provided by Eq.~(\ref{fullsoltk1}), and the term that describes the chameleon force inside body 
2 is replaced by
\begin{equation}
\label{fuerzaC1}
F_{\rm z\varphi_{\rm in2}} =-\frac{\partial}{\partial D}\int_{V_2}\varphi_{\rm in2}(\vec{r}^{\,\prime})\frac{\rho_2\beta}{2M_{pl}}d^3{\vec r}^{\,\prime}\;,
\end{equation}
where $\varphi_{\rm in2}$ is the solution given by Eq.(\ref{fullsoltk1}) for $0\leq r'\leq R_2$.

However, when both bodies are in the {\it thick shell regime}, only the expression $F_{\rm z\varphi_{\rm out}}$ remains the same as 
in Eq.(\ref{fuerzaC}) and the solution for $\varphi$ is provided by Eq.(\ref{fullsoltk2}) and $F_{\rm z\varphi_{\rm in2}}$ 
is given by Eq.~(\ref{fuerzaC1}) using Eq.~(\ref{fullsoltk2}) for $\varphi_{\rm in2}$, and $F_{\rm z\varphi_{\rm in1}}$ is replaced by 
\begin{equation}
\label{fuerzaC2}
F_{\rm z\varphi_{\rm in1}} =-\frac{\partial}{\partial D}\int_{V_1}\varphi_{\rm in1}(\vec{r})\frac{\rho_1\beta}{2M_{pl}}d^3\vec{r}\;,
\end{equation}
where the value of $\varphi_{\rm in1}$ is given by Eq.~(\ref{fullsoltk2}).

\section{WEP PREDICTIONS}
\label{WEPP}

In contrast with the conclusions obtained in the {\it standard approach}, and according to our analysis, for universal couplings
$\beta$,  and in the {\it thick shell regime}, the chameleon mediated force does depend  in a  relevant  manner on the composition of test bodies. This dependence is masked in Eqs.(\ref{fuerzaC}), (\ref{fuerzaC1}), (\ref{fuerzaC2}) but can be seen explicitly 
in the analytic expressions for the coefficients $C_l$ appearing in the  expansions (\ref{fullsoltk1}) and (\ref{fullsoltk2}) 
(cf. \cite{KLSSV18}).

In this section, we compute the theoretical prediction for the E\"otv\"os parameter $\eta$ in the LLR scenario 
and then compare our predictions under the three different kind of approximate methods discussed above. The parameter $\eta= 2\frac{|\vec{a_1}-\vec{a_2}|}{|\vec{a_1}+\vec{a_2}|}$ 
is associated with the differential acceleration of two bodies of different composition, where $\vec{a_i}=\vec{a}_{i\varphi}+\vec{g}$ ($i=1,2$) 
is the acceleration of the $i-$test body due to the combined chameleon force $\vec{F}_{i,\varphi}$ and the force of gravity $\vec{F}_{i,g}$, which is basically due to the large body. The acceleration $\vec{a}_{i\varphi}$ is found from the different expressions 
for the force presented in Sec.~\ref{sec:force}. It should be noted that in this paper we do not consider the case $M = 2.4 \times 10^{-3}$ {\rm eV}, which is associated with the {\it cosmological chameleon}
\footnote{The cosmological chameleon refers to the chameleon field that is responsible for the late time accelerated expansion of the Universe.}, because for this case, and under   the relevant conditions, one is almost always in the {\it thin shell regime}, and thus, our previous calculations remain valid.
  As shown in our previous estimations \cite{KLSSV18}, the predictions for the LLR experiment (when the bodies are in the {\it thin shell regime}) are very similar for both the {\it standard} and the two body approach and that their values are  below the experimental bound. As the largest value of $\eta$ corresponds to the case $n=1$ and $\beta=10^{-5}$ (in which the Moon looses its {\it thin shell} but the Earth and the Sun do not), we can be sure that the predictions for $\eta$ corresponding to  all the  {\it thick shell regime} are much smaller than the experimental bound.

\begin{figure*}
\subfloat[$n=1$] 
{\includegraphics[width=8.5cm,height=8.5cm]{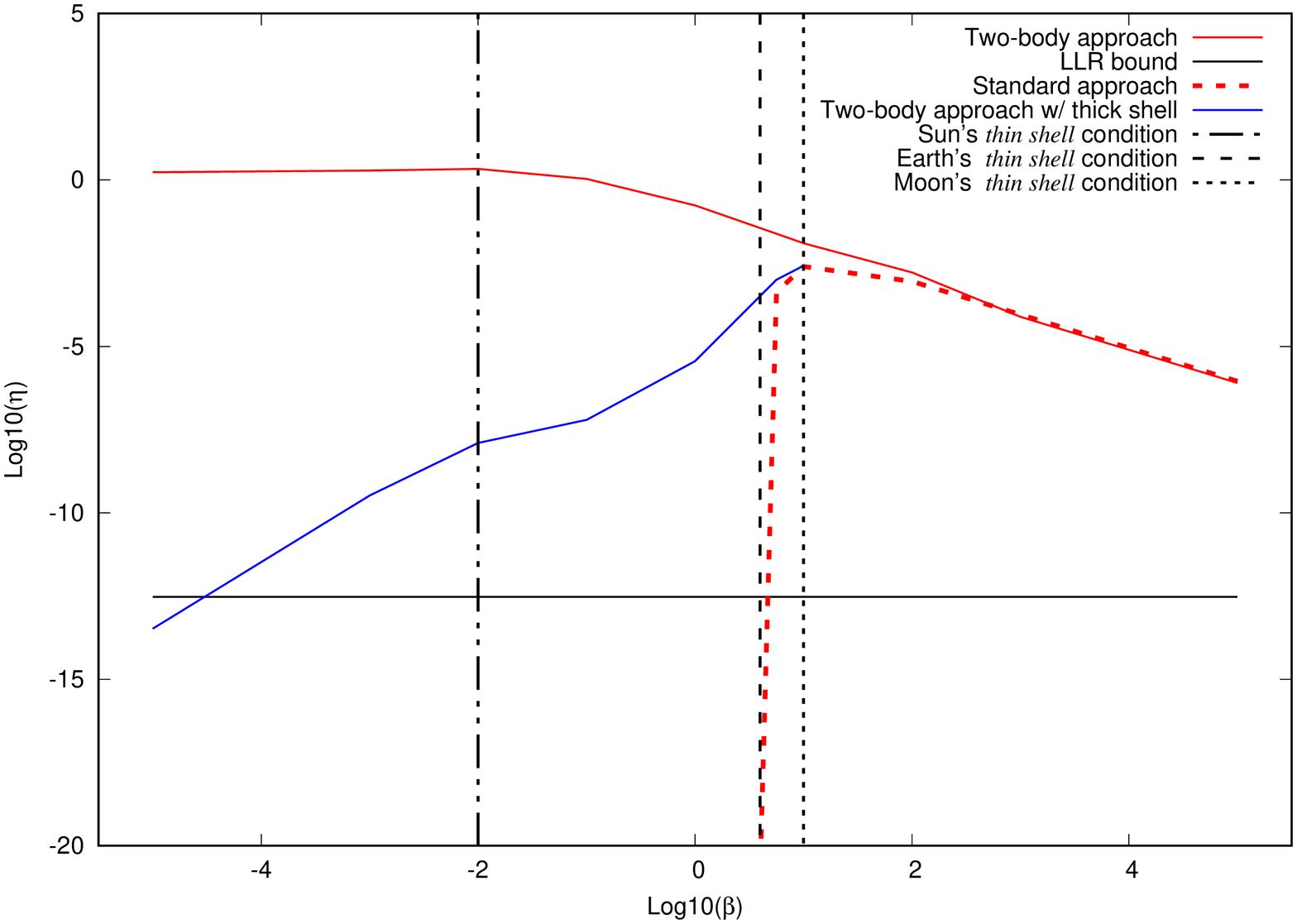}}
\subfloat[$n=2$]
{\includegraphics[width=8.5cm,height=8.5cm]{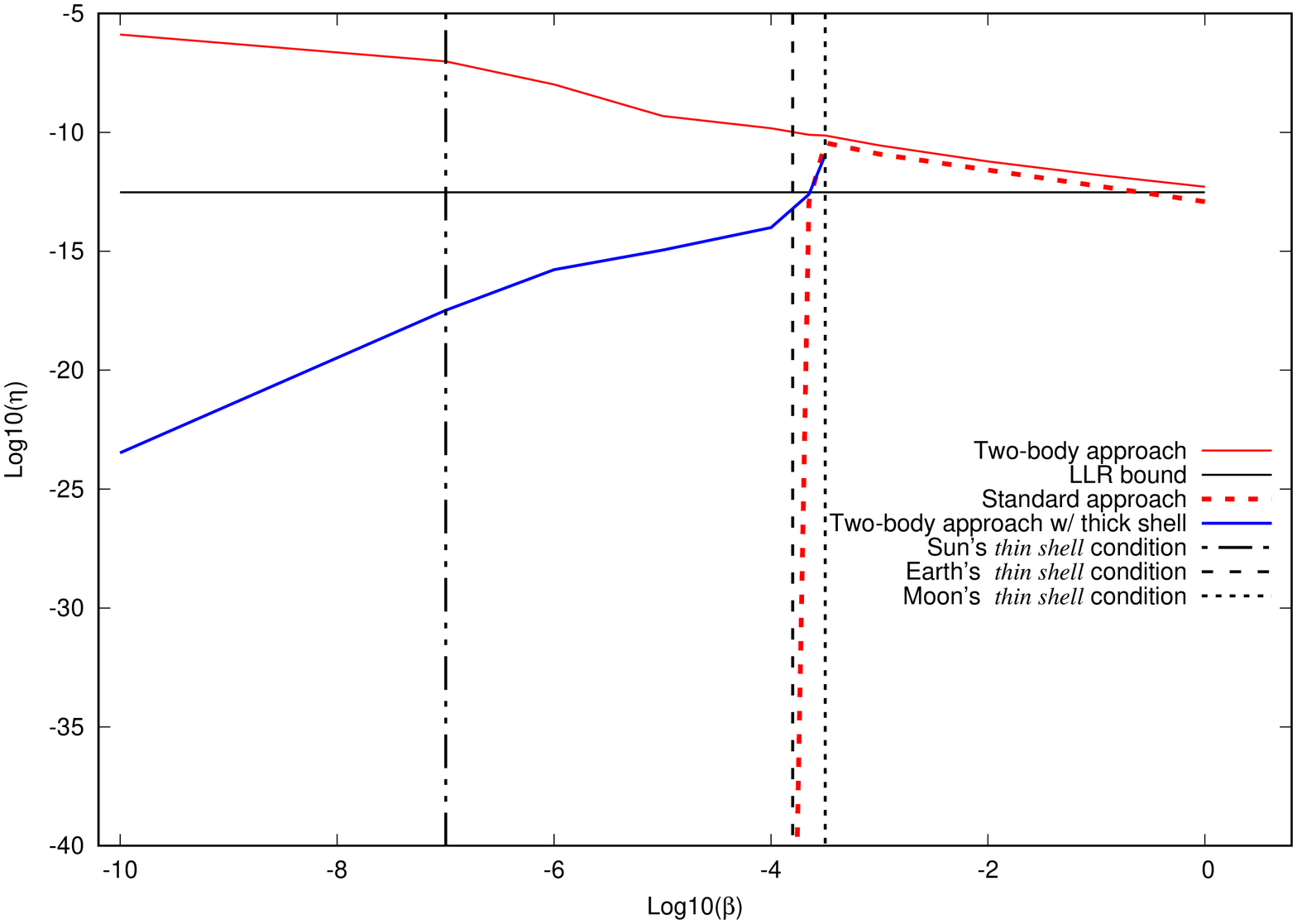}}
\caption{The E\"otv\"os parameter $\eta$ (in ${\rm log_{10}}$ scale) for the LLR experiment as a function of the parameter $\beta$ (in ${\rm log_{10}}$ scale) for different positive values of $n$. All bodies are surrounded by the interstellar medium. ${M = 10 \hspace{0.1cm} {\rm eV}}$. The vertical lines show the values of $\beta$ below which the {\it thin shell condition} is no longer satisfied for the Earth, Moon and Sun. 
The horizontal line represents the experimental bound~\cite{LLR}. For all values computed in this plot, the energy criterion developed in Section \ref{sec:energy} indicates that the two body approach together with an improved approximation to the effective potential 
provides better results than the {\it standard approach} and than the two-body approach with the quadratic approximation for the potential.} 
\label{resLLR3}
\end{figure*}

Figure \ref{resLLR3} depicts the predictions for $\eta$ based on the LLR experiment using three different approaches: {\it standard}, 
two-body with quadratic effective potential and the improved two-body {\it thick shell} approximation. 
 As can be seen  in the figure  
the largest value of the E\"otv\"os parameter is found when one of the test bodies is on the {\it thick shell regime}
but the other test body is not \cite{Hui2009}. This is in agreement with  the  conclusions of the {\it standard approach}. However,  in contrast  with the  latter, and  as    indicated in  Fig.\ref{resLLR3}  the acceleration generated by the chameleon force in test bodies which  are in the {\it thick shell regime} 
does depend on their composition (the same happens when the large body develops a {\it thick shell}). This  result is one of the most important outcomes  of this paper.
 Moreover, the predictions for $\eta$ decrease for lower values of $\beta$ but it does not become null when the {\it thin shell} 
condition does not ensue in the Earth, in contrast  with  what is predicted by the standard approach.  It should be noted that, even though, the predictions for $\eta$ ($n=1$) obtained in this paper are different from Ref.\cite{KLSSV18}, the conclusions are the same in that the chameleon model is ruled out for $\beta > 10^{-4}$. On the other hand, for $n=2$ 
the improved two body approach that implements the {\it thick shell} approximation is consistent with experimental bounds if $\beta < 10^{-3.8}$, contrary to what we found previously using the two body approach with the quadratic approximation for the effective potential.

\section{CONCLUSIONS}
\label{conclu}

 In this article we study the chameleon model using a two-body-problem approach devised
  by us in~\cite{KLSSV18}, by improving our previous analysis to situations where the {\it thick shell regime} becomes relevant.
In that regime we thus replace  the previously used
 quadratic approximation for the effective potential (which is not adequate when the field departs largely from its 
 minimum inside the bodies) by a linear one. This analysis amends our previous work \cite{KLSSV18} method in several respects.

First, we find expressions for the chameleon field $\varphi$ when one or both bodies (the test and the large bodies) 
are in the {\it thick shell regime}. Then, we calculate an energy functional and conclude that the  energy  minimization criterion favors our improved approach over the other ones   for  the  regimes in question.

We also obtain expressions for the chameleon mediated force from first principles and find predictions for the E\"ot\"vos parameter $\eta$ 
in a setup that mimics the LLR experiment. We conclude that our improved approach and the {\it standard approach} agrees on the predictions 
for $\eta$ when the chameleon coupling  lies in the following ranges 
$5\gtrsim\beta\gtrsim 10^{2.5}$ for $n=1$ and $1\gtrsim\beta\gtrsim 10^{-3.5}$ for $n=2$ with $M\sim 10$ {\rm eV} and
 the prediction is that the E\"otv\"os parameter 
is in conflict with the observational bounds imposed by the LLR experiments. The largest prediction of $\eta$ (for each $n$) corresponds to scenarios  where one of the test bodies is in the {\it thick shell regime} but the other is not. For lower  values of $\beta$ the {\it test bodies}  (the Moon and the Earth) and/or the large body (the Sun) are not in the {\it thin shell regime} anymore and thus our improved and the standard approaches do not agree. For instance, for $\beta\lesssim 10$ ($n=1$) the {\it standard approach}
predicts no violation of the experimental bounds for $\eta$, while our improved treatment shows that for $10^{-4}\lesssim \beta \lesssim 10^4$ 
the predicted $\eta$ violates the experimental bounds. Our results allow us then to put further constraints on the parameters 
$n,\beta,M$ of the original chameleon model. Finally, we stress that   in contrast  with the conclusions obtained from the {\it standard approach}, our treatment shows that test bodies 
having a {\it thick shell} fall with accelerations that are composition dependent. 
The current and previous analyses \cite{KLSSV18} illustrate the difficulty in considering a sharp 
distinction  between the {\it thin} and {\it thick shell regimes}. In fact, whenever an approximation is used relying on one or the other regimes 
or the consideration of the one body problem (standard approach) {\it versus} the two body problem (our approach), 
there is no {\it a priori} manner to be sure which one is more appropriate at an arbitrary level of precision. In this 
regard the energy criteria seems to offer a reliable guidance as to which approximation is more trustworthy.

\section*{Acknowledgements}
The authors acknowledge the use of the supercluster MIZTLI  of UNAM through project LANCAD-UNAM-DGTIC-132 and thank the people of DGTIC-UNAM for technical and computational support.
The authors thank Carolina Negrelli for help with the numerical calculations. L.K. and S.L. are supported  by CONICET Grant No. PIP 11220120100504 and by the National Agency for the Promotion of Science and Technology (ANPCYT) of Argentina Grant No. PICT-2016-0081; and with H.V. by Grant No. G140 from UNLP.  M.S. is partially supported by UNAM-PAPIIT Grant No.s IN107113, IN111719 and CONACYT Grant No. CB-166656. D.S. is supported in part by CONACYT No. 101712,  and PAPIIT- UNAM No. IG100316 M\'{e}xico, as well as  sabbatical  fellowships  from  PASPA-DGAPA-UNAM-M\'{e}xico, and   from  Fulbright-Garcia Robles-COMEXUS.

\bibliographystyle{apsrev}
\bibliography{chamthickshell2019b}

\end{document}